\documentclass[traditabstract]{aa}

\usepackage{natbib}
\usepackage{graphicx}
\usepackage{amssymb}
\usepackage[dvipsnames]{color}

\begin{document}

\newcommand{\3}{\ss}
\newcommand{\n}{\noindent}
\newcommand{\eps}{\varepsilon}
\newcommand{\be}{\begin{equation}}
\newcommand{\ee}{\end{equation}}
\newcommand{\bl}[1]{\mbox{\boldmath$ #1 $}}
\def\ba{\begin{eqnarray}}
\def\ea{\end{eqnarray}}
\def\de{\partial}
\def\msun{M_\odot}
\def\div{\nabla\cdot}
\def\grad{\nabla}
\def\rot{\nabla\times}
\def\ltsima{$\; \buildrel < \over \sim \;$}
\def\simlt{\lower.5ex\hbox{\ltsima}}
\def\gtsima{$\; \buildrel > \over \sim \;$}
\def\simgt{\lower.5ex\hbox{\gtsima}}

\title{The effect of episodic accretion on the phase transition of 
CO and CO$_2$ in low-mass star formation}
\titlerunning{Episodic accretion and the abundance of ices}

\author{Eduard~I.~Vorobyov \inst{1,2}, Isabelle Baraffe \inst{3}, Tim Harries \inst{3}, Gilles Chabrier \inst{4,3} }
\authorrunning{Vorobyov et al. }

\offprints{E.~I.~Vorobyov} 
   
\institute{University of Vienna, Institute of Astrophysics,  Vienna, 1180, Austria; 
\email{eduard.vorobiev@univie.ac.at}
\and
Research Institute of Physics, Southern Federal University, Stachki Ave. 194, Rostov-on-Don, 
344090 Russia 
\and 
University of Exeter, Physics and Astronomy, Stocker Road, EX4 4QL Exeter
\email{i.baraffe@ex.ac.uk, T.J.Harries@exeter.ac.uk}
\and
\'Ecole Normale Sup\'erieure, Lyon, CRAL (UMR CNRS 5574), Universit\'e de Lyon, France
\email{gilles.chabrier@ens-lyon.fr}
}

\date{}

\abstract{
We study the evaporation and condensation of CO and CO$_2$ during the
embedded stages of low-mass star formation by using numerical simulations. 
We focus on the effect of luminosity
bursts, similar in magnitude
to FUors and EXors, on the gas-phase abundance of CO and CO$_2$ in the
protostellar disk and infalling
envelope. 
The evolution of a young protostar and its environment is followed based on hydrodynamical models using the thin-disk approximation, coupled
with a stellar evolution
code and phase transformations of CO and CO$_2$. The accretion and associated
luminosity bursts in our model
are caused by disk gravitational fragmentation followed by quick migration
of the fragments onto the forming protostar.
We found that bursts with luminosity on the order of 100--200$~L_\odot$
can evaporate CO ices in part of the envelope. The typical freeze-out
time of the gas-phase CO onto dust grains in the envelope (a few kyr) is
much longer than
the burst duration (100--200~yr). This results in an increased abundance
of the gas-phase CO in the envelope long after the system has returned
into a quiescent stage.
In contrast, luminosity bursts can evaporate CO$_2$ ices only in the
disk, where the
freeze-out time of the gas-phase CO$_2$ is comparable to the burst duration.
We thus confirm that luminosity bursts can leave long-lasting traces in the
abundance of
gas-phase CO in the infalling envelope, enabling the detection of recent
bursts
as suggested by previous semi-analytical studies. }


\maketitle

\section{Introduction}

Low-mass stars are born as a result of the gravitational collapse of dense and cold 
cores composed mainly of molecular hydrogen, which in turn are formed in giant molecular clouds 
from the combination of turbulent, gravitational and magnetic processes \citep[e.g][]{mckee12}. 
Despite many observational and theoretical efforts, how a low-mass  star accumulates its final mass remains an open question. In the classic model of inside-out collapse \citep{Shu77}, the mass accretion 
rate onto the protostar is proportional to the cube of the sound speed, 
$\dot{M} \propto c_{\rm s}^3 /G$, implying 
a rather narrow spread of accretion rates  $(2-5)\times 10^{-6}~M_\odot$~yr$^{-1}$
for typical conditions in pre-stellar cores. These values of $\dot{M}$ are in stark
disagreement with those inferred for young protostars, exhibiting 
a variety of accretion rates from  $\la 10^{-7}~M_\odot$~yr$^{-1}$ to 
$\ga 10^{-5}~M_\odot$~yr$^{-1}$ \citep{Dunham06,Evans09}. 
In particular, the fraction of objects with $\dot{M}<10^{-6}~M_\odot$~yr$^{-1}$
amounts to 50\% in Perseus, Serpens, and Ophiuchus star-forming regions \citep{Enoch09}.
FU-Orionis-type objects (FUors) with accretion rates sometimes exceeding
$10^{-4}~M_\odot$~yr$^{-1}$ and very-low-luminosity objects \citep[e.g][]{Bourke06} with accretion
rates most certainly below $10^{-6}~M_\odot$~yr$^{-1}$ also do not fit into the classic model.

It is becoming evident that the simplified model of \citet{Shu77} cannot explain the whole variety 
of inferred accretion rates and the concept of variable accretion with episodic bursts, 
based on the original idea
of \citet{Kenyon90}, is currently gaining theoretical and observational support.
According to this concept, the mass accretion onto the forming protostar  is characterized
by short ($\la 100-200$~yr) bursts approximately  
with $\dot{M} \ga 10^{-5}~M_\odot$~yr$^{-1}$ alternated with 
longer ($ 10^{3}$--$10^{4}$~yr) quiescent periods with $\dot{M} \la 10^{-6}~M_\odot$~yr$^{-1}$.

Numerous mechanisms that can produce variable accretion with episodic
bursts have been proposed in the past. They include
models explaining the origin of FUor accretion bursts by means of 
viscous-thermal instabilities in the inner disk \citep{Lin86,Bell94}, 
thermal instabilities induced by density perturbations due to, e.g., a massive planet in the disk 
\citep{LC04}, and tidal effects from close encounters  in binary systems or stellar clusters
\citep{BB92,Pfalzner08}. 
Recently, several promising accretion burst mechanisms have emerged or received further 
refinement, such as a combination of gravitational instability and the triggering of the 
magnetorotational instability \citep{Armitage01,Zhu10}, 
accretion of dense clumps in a gravitationally fragmenting disk \citep{VB06,VB10}, 
perturbations caused by a planetary or sub-stellar object on an eccentric orbit 
\citep{Machida2011,Vor2012}, 
or instability when an accretion disk is truncated by the star's strong magnetic field close to 
the corotation radius \citep{DAS10}.

Episodic accretion may have important consequences for the evolution of stars and planets. 
It provides an explanation for the long-standing luminosity problem in young protostars \citep{DV12},
explains the luminosity spread in young stellar clusters without invoking any significant age 
spread \citep{Baraffe09,Baraffe12}, and can account for unexpected lithium and beryllium 
depletion in some young stars \citep{Baraffe10,Viallet12}.
In addition, quiescent periods between the accretion bursts can promote disk fragmentation, 
which might in turn promote planet formation \citep{Stamatellos11}.

Episodic accretion may also have important implications for 
the chemical evolution of protostellar disks and envelopes. Using simplified core collapse 
calculations with prescribed accretion bursts, \citet{Lee07} and \citet{Visser12} 
showed that the bursts can lead to evaporation of CO, CO$_2$, and some other ices in protostellar envelopes.
The freeze-out timescale of these species is expected to be longer than the typical duration of the
burst ($\sim 100$~yr), so that the chemical signatures of the bursts can linger through the quiescent
phase of accretion. \citet{Kim11,Kim12} found evidence for pure CO$_2$ ice  in about 50\% 
of the observed low-luminosity
sources, implying that the dust temperature must have been higher in the past presumably due to 
recent accretion bursts. 

In this paper, we employ numerical hydrodynamics simulations coupled with a stellar evolution 
code, as done in \cite{Baraffe12},  and a simplified chemical model, 
to explore the effect of episodic accretion onto the evaporation and freeze-out of CO and CO$_2$ during the early stages of star formation. 
The coupling between hydrodynamical simulations and the evolution of the central, accreting object provides a consistent value of the protostar luminosity, which is relevant for the estimate of the radiative feedback from the protostar in the disk and the envelope. This feedback impacts the gas/dust temperatures which are required for the calculations of CO and CO$_2$ abundances. In the present hydrodynamical calculations, estimate of the gas/dust temperatures relies on the diffusion approximation for the treatment of radiative transfer and additional approximations regarding the geometry of the disk and of the surrounding 
material. 
We analyse the uncertainties of temperatures  based on such approximations, by comparing them with temperatures obtained from improved   radiative transfer calculations based on  the TORUS code (Harries et al. 2004; Harries 2011). We will also discuss (\S 3) the impact of our approximation on the predicted CO and CO$_2$ abundances.

\section{Model description}
\subsection{Basic equations}
\label{model}
Our numerical hydrodynamics model for the formation and evolution of a young stellar object 
is described in detail in \citet{VB10}. Here, we briefly review the main concepts and describe
latest modifications. The model in the most general case includes
a forming protostar, described by a stellar evolution code, and a protostellar disk 
plus infalling envelope, both described by numerical hydrodynamics equations.
We use the thin-disk approximation, which is an excellent 
means to calculate the evolution for many orbital periods and many model parameters
and its justification is discussed in \citet{VB10}. 
The thin-disk approximation is complemented by a calculation of the vertical scale height $h$
in both the disk and envelope determined in each computational cell using an assumption of local hydrostatic
equilibrium. The resulting model has a flared structure with the vertical
scale height increasing with radial distance according to the law $h\propto r^{1.5}$.
Both the disk and envelope receive a fraction of the irradiation energy 
from the central protostar described by equation~(\ref{fluxF}) (see below). 
The main physical processes taken into account when computing the evolution of the 
disk and envelope include viscous and shock heating, irradiation by the forming star, 
background irradiation, radiative cooling from the disk surface and self-gravity. 
The corresponding equations of mass, momentum, and energy transport  are
\begin{equation}
\label{cont}
\frac{{\partial \Sigma }}{{\partial t}} =  - \nabla_p  \cdot 
\left( \Sigma \bl{v}_p \right),  
\end{equation}
\begin{eqnarray}
\label{mom}
\frac{\partial}{\partial t} \left( \Sigma \bl{v}_p \right) &+& \left[ \nabla \cdot \left( \Sigma \bl{v_p}
\otimes \bl{v}_p \right) \right]_p =   - \nabla_p {\cal P}  + \Sigma \, \bl{g}_p + \\ \nonumber
& + & (\nabla \cdot \mathbf{\Pi})_p, 
\label{energ}
\end{eqnarray}
\begin{equation}
\frac{\partial e}{\partial t} +\nabla_p \cdot \left( e \bl{v}_p \right) = -{\cal P} 
(\nabla_p \cdot \bl{v}_{p}) -\Lambda +\Gamma + 
\left(\nabla \bl{v}\right)_{pp^\prime}:\Pi_{pp^\prime}, 
\end{equation}
where subscripts $p$ and $p^\prime$ refers to the planar components $(r,\phi)$ 
in polar coordinates, $\Sigma$ is the mass surface density, $e$ is the internal energy per 
surface area, 
${\cal P}$ is the vertically integrated gas pressure calculated via the ideal equation of state 
as ${\cal P}=(\gamma-1) e$ with $\gamma=7/5$,
$\bl{v}_{p}=v_r \hat{\bl r}+ v_\phi \hat{\bl \phi}$ is the velocity in the
disk plane, and $\nabla_p=\hat{\bl r} \partial / \partial r + \hat{\bl \phi} r^{-1} 
\partial / \partial \phi $ is the gradient along the planar coordinates of the disk. 
The gravitational acceleration in the disk plane, $\bl{g}_{p}=g_r \hat{\bl r} +g_\phi \hat{\bl \phi}$, takes into account self-gravity of the disk, found by solving for the Poisson integral 
\citep[see details in][]{VB10}, and the gravity of the central protostar when formed. 

Turbulent viscosity due to sources other than gravity 
is taken into account via the viscous stress tensor 
$\mathbf{\Pi}$, the expression for which is provided in \citet{VB10}.
We parameterize the magnitude of kinematic viscosity $\nu$ using the $\alpha$-prescription 
with a spatially and temporally uniform $\alpha=5\times 10^{-3}$.

The radiative cooling $\Lambda$ in equation~(\ref{energ}) is determined using the diffusion
approximation of the vertical radiation transport in a one-zone model of the vertical disk 
structure \citep{Johnson03}
\begin{equation}
\Lambda={\cal F}_{\rm c}\sigma\, T_{\rm mp}^4 \frac{\tau}{1+\tau^2},
\end{equation}
where $\sigma$ is the Stefan-Boltzmann constant, $T_{\rm mp}={\cal P} \mu / R \Sigma$ is 
the midplane temperature of gas\footnote{This definition of the midplane temperature is accurate within
a factor of order unity \citep{Zhu2012}}, $\mu=2.33$ is the mean molecular weight, 
$R$ is the universal 
gas constant, and ${\cal F}_{\rm c}=2+20\tan^{-1}(\tau)/(3\pi)$ is a function that 
secures a correct transition between the optically thick and optically thin regimes.  
We use frequency-integrated opacities of \citet{Bell94}.
The heating function is expressed as
\begin{equation}
\Gamma={\cal F}_{\rm c}\sigma\, T_{\rm irr}^4 \frac{\tau}{1+\tau^2},
\end{equation}
where $T_{\rm irr}$ is the irradiation temperature at the disk surface 
determined by the stellar and background black-body irradiation as
\begin{equation}
T_{\rm irr}^4=T_{\rm bg}^4+\frac{F_{\rm irr}(r)}{\sigma},
\label{fluxCS}
\end{equation}
where $T_{\rm bg}$ is the uniform background temperature (in our model set to the 
initial temperature of the natal cloud core)
and $F_{\rm irr}(r)$ is the radiation flux (energy per unit time per unit surface area) 
absorbed by the disk surface at radial distance 
$r$ from the central star. The latter quantity is calculated as 
\begin{equation}
F_{\rm irr}(r)= \frac{L_\ast}{4\pi r^2} \cos{\gamma_{\rm irr}},
\label{fluxF}
\end{equation}
where $\gamma_{\rm irr}$ is the incidence angle of 
radiation arriving at the disk surface at radial distance $r$. The incidence angle is calculated
using the disk surface curvature inferred from the radial profile of the  
disk vertical scale height \citep[see][for more details]{VB10}.

The central object's luminosity $L_\ast$ is the sum of the accretion luminosity $L_{\rm \ast,accr}=G M_\ast \dot{M}/2
R_\ast$ arising from the gravitational energy of accreted gas and
the photospheric luminosity $L_{\rm \ast,ph}$ due to gravitational contraction and deuterium burning
in the protostar interior. The stellar mass $M_\ast$ and accretion rate onto the star $\dot{M}$
are determined self-consistently during numerical simulations using the amount of gas passing through
the sink cell. 
The evolution of the accreting protostar is based on the Lyon stellar evolution code with input physics described in \cite{Chabrier97} and including accretion processes as described in  \citet{Baraffe09} and \citet{Baraffe12}. The accretion rates are derived from the hydrodynamic calculations above described. As in \citet{Baraffe12}, we assume that a fraction $\alpha$ of the accretion energy  
$\epsilon\frac{G M_\ast{\dot M}}{R_\ast}$ is absorbed by the protostar, while the fraction (1-$\alpha$) is radiated away and contributes to $L_\ast$ in equation (\ref{fluxF})\footnote{As in \citet{Baraffe09}, we assume a value $\epsilon$=1/2 characteristic of accretion from a thin disk.}. In the present calculations, we adopt a "hybrid" scheme to describe the contribution of the accreted matter to the protostar's internal energy, with "cold" accretion, i.e $\alpha=0$, when accretion rates remain smaller than  a critical value 
$\dot{M}_{\rm cr}$, and  "hot" accretion,  i.e $\alpha \ne 0$,  when $\dot{M} >  \dot{M}_{\rm cr}$. In this paper we adopt $\dot{M}_{\rm cr}=10^{-5}~M_\odot$~yr$^{-1}$ and $\alpha=0.2$ \citep[see discussion in ][]{Baraffe12}.  For the initial mass of the protostar,  corresponding to the second Larson core mass,  we adopt a value of 1.0$~M_{\rm
Jup}$ with an initial radius  $\sim 1.0~R_\odot$.

The input parameters provided by the hydrodynamic calculations to the coupled stellar evolution code are the age of the protostar and the accretion rate 
onto the stellar surface $\dot{M}$, 
while the output are the radius and the photospheric luminosity of the protostar. 
The stellar evolution code is called to update the properties of the protostar 
every 5~yr of the physical time. For comparison, the global hydrodynamical timestep may be as small as
a few weeks and the whole cycle of numerical simulations may exceed 1.0 Myr.
This coupling of the disk and protostar evolution allows for a self-consistent determination
of the radiative input of the protostar into the disk thermal balance, which is important for
the accurate study of disk instability and fragmentation.

The dynamics of carbon monoxide (CO) and carbon dioxide (CO$_2$) including absorption onto 
and desorption from dust grains is computed using the modified equations of 
continuity for the surface densities of solid ($\Sigma^{\rm s}_{\rm i}$) and gaseous 
($\Sigma^{\rm g}_{\rm i}$) phases 
\begin{eqnarray}
\label{COgas}
{\partial \Sigma^{\rm g}_{\rm i} \over \partial t} &+& \nabla_{\rm p} \cdot 
\left(\Sigma^{\rm g}_{\rm i} {\bl v}  \right) = -\lambda 
\Sigma^{\rm g}_{\rm i} + \eta \\
\label{COsolid}
{\partial \Sigma^{\rm s}_{\rm i} \over \partial t} &+& \nabla_{\rm p} \cdot 
\left( \Sigma^{\rm s}_{\rm i} {\bl v} \right)
= \lambda \Sigma^{\rm g}_{\rm i} -\eta,
\end{eqnarray}
where index $i$ corresponds to either CO or CO$_2$, $\lambda$ is the absorption rate (s$^{-1}$)
and $\eta$ is the desorption velocity from dust grains in units of g~cm$^{-2}$~s$^{-1}$. The expressions
for $\eta$ and $\lambda$ were taken
from \citet{Charnley01} and \citet{Visser09} assuming a zero-order desorption for thick ice mantles
\begin{equation}
\label{lambda}
\lambda = 1.45\times 10^4 \left( {T_{\rm mp} \over M_{\rm i}} \right)^{0.5} \langle \pi a^2 n_{\rm d}  \rangle,
\end{equation}
\begin{equation}
\eta = 4 \pi a^2 n_{\rm d} \nu_{\rm i} \exp{\left[-{E_{\rm i} \over k T_{\rm d}}\right]} 2 h m_{\rm i},
\end{equation} 
where $T_{\rm mp}$ is the gas midplane temperature\footnote{We neglect possible vertical variations
in the gas temperature. A more accurate approach involving the reconstruction of the vertical structure
is currently in development.},  
$M_{\rm i}$ and  $m_{\rm i}$ are the molecular weight and mass (in gram), respectively,
and $n_{\rm d}$ is the number density of dust grains. 
The values of the binding energy for CO and CO$_2$ ($E_{\rm co}/k=885$~Ê and  $E_{\rm co_2}/k=2300$~K)
and vibrational frequency ($\nu_{\rm co}=7\times 10^{26}$~cm$^{-2}$~s$^{-1}$ and 
$\nu_{\rm co_2}=9.5\times 10^{26}$~cm$^{-2}$~s$^{-1}$) were taken from \citet{Bisschop06} and \citet{Noble12}.
In this paper, we make a simplifying assumption of $T_{\rm d}=T_{\rm mp}$ and discuss possible consequences
in Section~\ref{caveats}.

In the limit of a constant radius of dust grains $a$, the expression in brackets 
in equation~(\ref{lambda}) can be written in the following simplified form
\begin{equation}
\langle \pi a^2 n_{\rm d} \rangle =  {\pi a^2 A_{\rm d2g} \Sigma_{\rm g} \over 2 h m_{\rm d}},
\end{equation} 
where $A_{\rm d2g}=0.01$ is the dust to gas mass ratio, $m_{\rm d} = 4 \pi a^3 \rho_{d.p.}/3$ is the
mass of a dust grain and $\rho_{\rm d.p.}=2.5$~g~cm$^{-3}$ is the density of dust grains. 
The radius of dust grains is set in this study to $a=0.1~\mu$m.

When writing equations~(\ref{COgas}) and (\ref{COsolid}) for the surface densities of CO and CO$_2$
we assumed that dust is passively transported with gas and neglected the dust radial drift caused 
by the dust--gas drag force. The latter simplification is of little consequence
for the dynamics of dust grains with sizes $\lesssim 10$~$\mu m$ on
timescales of $\le 0.5$~Myr \citep{Takeuchi02}. We also do not take into account chemical 
reactions that may change the abundance of CO and CO$_2$, and consider only phase transformations
of these species. Chemical transformations will be taken into account in a follow-up study.

We start our numerical simulations from the gravitational collapse of a {\it starless} cloud core, 
continue into the embedded phase of star formation, during which
a star, disk, and envelope are formed, and terminate our simulations in the T Tauri phase,
when most of the envelope has accreted onto the forming star plus disk system.
The protostellar disk, when present, is located in the inner part of the numerical 
polar grid, while the collapsing envelope occupies the rest of the grid. 
As a result, the disk
is not isolated but is exposed to intense mass loading from the envelope.  
In addition, the mass accretion rate onto the disk is not a free parameter of the model 
but is self-consistently determined by the gas dynamics in the envelope.

To avoid too small time steps, we introduce a ``sink cell'' at $r_{\rm sc}=5$~AU and 
impose a free inflow inner boundary condition
and free outflow outer boundary condition so that the matter is allowed to flow out of 
the computational domain but is prevented from flowing in. 
The sink cell is dynamically inactive; it contributes only to the total gravitational 
potential and secures a smooth behaviour of the gravity force down to the stellar surface.
During the early stages of the core collapse, we monitor the gas surface density in 
the sink cell and when its value exceeds a critical value for the transition from 
isothermal to adiabatic evolution, we introduce a central point-mass object.
In the subsequent evolution, 90\% of the gas that crosses the inner boundary 
is assumed to land onto the protostar. A small fraction of this mass (a few per cent) 
remains in the sink cell
to guarantee a smooth transition of the gas surface density across the inner boundary. 
The other 10\% of the accreted gas is assumed to be carried away with protostellar jets. 

The numerical resolution is $512\times 512$ grid points and the numerical procedure to solve
hydrodynamics equations~(\ref{cont})-(\ref{energ})  is described in detail in \citet{VB10}. 
The modified continuity equations~(\ref{COgas}) and (\ref{COsolid}) are solved using a two-step procedure.
First, the advection part (zero r.h.s.) is solved using the same piecewise parabolic advection
scheme as for the gas surface density. Then, the surface densities of CO and CO$_2$ are updated
to take into account the phase transformations using a first-order backward Euler scheme. 
This scheme is implicit and in principle does not require a time-step limiter. However, too
big timesteps can result in the loss of accuracy. Therefore, if the relative change 
of $\Sigma^{\rm s}_{\rm i}$ and/or  
$\Sigma^{\rm g}_{\rm i}$ exceeds 10\% over one global hydro step, the local integration timestep
for equations~(\ref{COgas}) and (\ref{COsolid}) is reduced by a factor of 2 and the solution 
is sought once again. This subcycling procedure is repeated until the desired accuracy is achieved.

\subsection{Initial setup}
For the gas surface density $\Sigma$  and angular velocity $\Omega$ radial distributions
we take those typical of pre-stellar cores formed as a result of the slow expulsion 
of magnetic field due to ambipolar diffusion, with the angular
momentum remaining constant during axially-symmetric core compression \citep{Basu97}
\begin{equation}
\Sigma={r_0 \Sigma_0 \over \sqrt{r^2+r_0^2}}\:,
\label{dens}
\end{equation}
\begin{equation}
\Omega=2\Omega_0 \left( {r_0\over r}\right)^2 \left[\sqrt{1+\left({r\over r_0}\right)^2
} -1\right].
\label{omega}
\end{equation}
Here, $\Omega_0$ and $\Sigma_0$ are the angular velocity and gas surface
density at the center of the core and $r_0 =\sqrt{A} c_{\rm s}^2/\pi G \Sigma_0 $
is the radius of the central plateau, where $c_{\rm s}$ is the initial sound speed in the core. 
The gas surface density distribution described by equation~(\ref{dens}) can
be obtained (to within a factor of order unity) by integrating the 
three-dimensional gas density distribution characteristic of 
Bonnor-Ebert spheres with a positive density-perturbation amplitude A \citep{Dapp09}.
The value of $A$ is set to 1.2 and the initial gas temperature is set to 10~K.

In order to form a gravitationally unstable core, we set the ratio of the outer radius $r_{\rm out}$
to the radius of the central plateau $r_0$ to 6.0. For the outer radius of the core
$r_{\rm out}=16000$~AU, the resulting central surface density is 
$\Sigma_0=4.5\times 10^{-2}$~g~cm$^{-2}$ and the total mass of the core is $M_{\rm c}=1.23~M_\odot$.
The central angular velocity $\Omega_0$ is set to 1.0~km~s$^{-1}$~pc$^{-1}$, which yields the ratio
of rotational to gravitational energy $\beta=5\times10^{-3}$. Our previous numerical simulations 
indicate that pre-stellar cores with similar $M_{\rm c}$ and $\beta$ produce disks that are gravitationally
unstable. Mass accretion rates onto the protostar
in these models often exhibit episodic accretion bursts caused by disk fragmentation and migration 
of the fragments onto the protostar \citep{VB10}. 

The abundances of CO and CO$_2$ (relative to the number density of molecular hydrogen) are
set to  $5\times 10^{-5}$, typical for pre-stellar gravitationally unstable cores\footnote{The adopted abundances are not expected to influence our results because
we do not take into account chemical transformations in this study.} \citep{Charnley01,Pontoppidan08}.
We assume that initially 10\% of CO and CO$_2$ is in the solid phase in order to be 
consistent with previous works of  \citet{Visser09} and 
\citet{Visser12}. 

\begin{figure*}
  \centering
  \includegraphics[width=15cm]{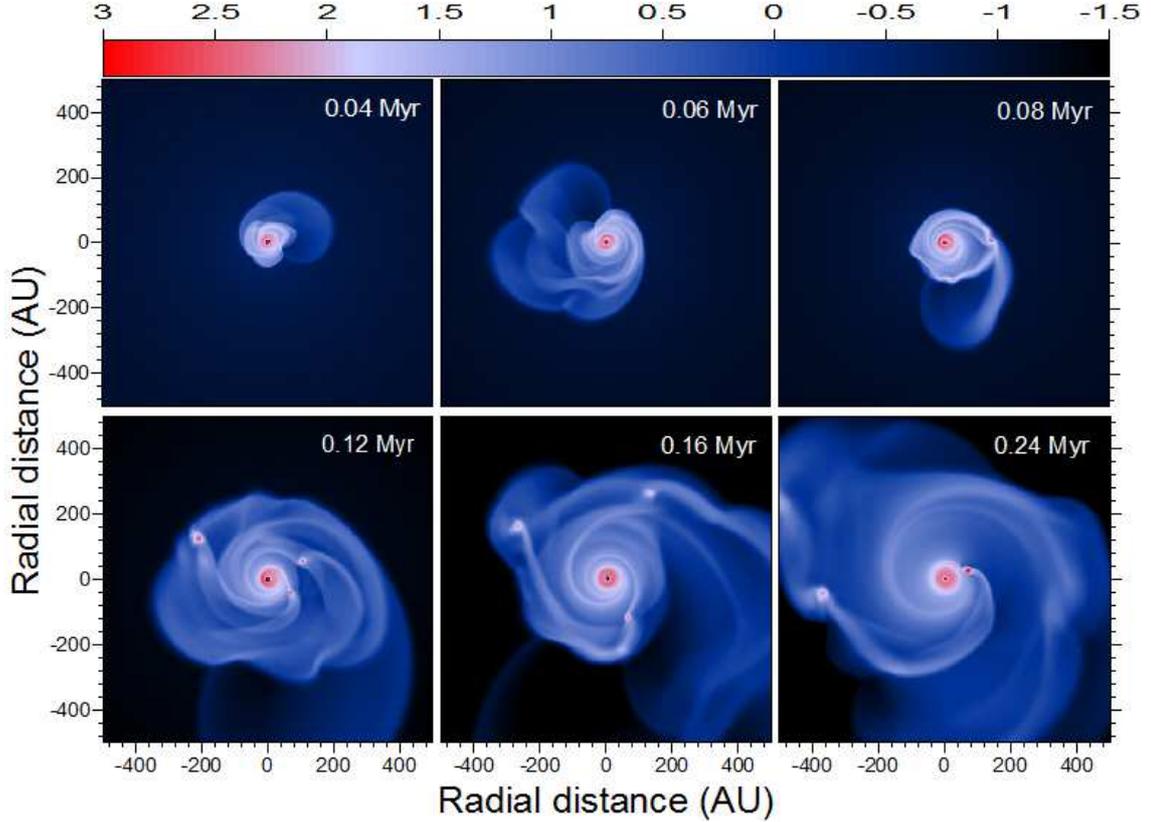}
  \caption{Gas surface density images in the inner $1000\times 1000$ AU showing the disk
  evolution during 0.24~Myr after the formation of the central star. Vigorous disk fragmentation
  is evident in the figure. The scale bar is in 
  log~g~cm$^{-2}$.}
  \label{fig1}
\end{figure*}

\section{Results of numerical simulations}
\subsection{Main characteristics of the model}
We start by describing the main properties of the forming protostar, disk and envelope. 
Figure~\ref{fig1} shows the gas surface density maps in the inner $1000 \times 1000$~AU box 
at six consecutive times after formation of the central protostar. The disk forms at $t=0.01$~Myr,
grows in mass and size due to continuing mass loading from the infalling envelope, and becomes gravitationally
unstable (as manifested by a weak spiral structure) as early as at $t=0.04$~Myr. 
The first episode of disk fragmentation occurs at $t\approx0.08$~Myr. In the subsequent evolution,
multiple fragments emerge in the disk at distances $\ga 50$~AU but most migrate into the inner regions and through the sink cell due to the loss of angular momentum via gravitational interaction with
the spiral arms and other fragments in the disk. This  phenomenon is studied in detail by \citet{VB06,VB10} and is confirmed by independent
fully three-dimensional studies \citep[e.g.][]{Machida2011}.
The ultimate fate of the migrating fragments depends on how quickly they can 
contract to stellar or planetary-sized objects. If the contraction timescale is longer 
than the migration/tidal destruction timescale, then 
the fragments will be completely destroyed when approaching the protostar, releasing 
their gravitational energy in the form of luminosity outbursts comparable to magnitude to 
FU Orionis and EX Lupi objects. This scenario is assumed in the present work.

\begin{figure}
  \resizebox{\hsize}{!}{\includegraphics{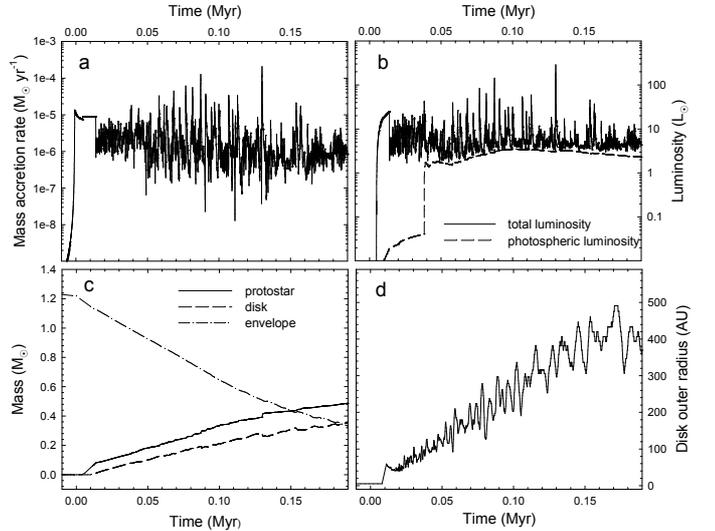}}
  \caption{Main characteristics of the forming system as a function of time elapsed
  since the formation of the central star: {\bf a)} the mass accretion rate 
  onto the protostar, {\bf b)} the total and photospehric luminosities, {\bf c)} the masses
  of the protostar, disk, and envelope, and {\bf d)} the disk radius. }
  \label{fig2}
\end{figure}

Figure~\ref{fig2} presents the time evolution of 
{\bf a)} the mass accretion rate onto the star $\dot{M}$, 
{\bf b)} the photospheric (dashed line) and total (solid line) luminosities, {\bf c)}
the masses of the protostar (solid line), protostellar disk (dashed line), 
and envelope (dash-dotted line), and
{\bf d)} the radius of the disk. The partition between the disk and infalling envelope 
is based on a threshold density of $\Sigma_{d2e}=0.5$~g~cm$^{-2}$ 
and the corresponding algorithm is described in detail in \citet{DV12}.
The time is counted from the formation of the central protostar.

Two luminosity outbursts exceeding in magnitude $100~L_\odot$, along with a number of 
smaller bursts ($\ga \mathrm{several}\times 10~L_\odot$),  are evident 
against the background luminosity of a few $L_\odot$. The photospheric luminosity provides a 
negligible input into the total flux in the early evolution ($t<0.04$~Myr) because 
the accretion rate does not exceed $10^{-5}~M_\odot$~yr$^{-1}$ and the accretion 
process is essentially cold, depositing little entropy to the protostar
The first episode of hot accretion occurs at $t\approx 0.04$~Myr, when the
mass accretion rate exceeds $10^{-5}~M_\odot$~yr~$^{-1}$. Thereafter,
the photospheric luminosity is slowly varying around $2-3~L_\odot$.

Panels {\bf c)} and {\bf d)} in Figure~\ref{fig2} illustrate the time evolution of other 
global characteristics of our model. The disk mass steadily grows during the initial 
evolution and reaches a value of $M_{\rm d} \approx 0.35~M_\odot$ at 0.2~Myr. 
In the subsequent evolution, the disk mass saturates and starts to decline slowly. 
The mass of the protostar 
exceeds that of the disk and shows episodic sharp increases caused by massive fragments 
spiralling in onto the protostar. The envelope mass steadily decreases and the model enters the
class I stage of stellar evolution, defined as the time when approximately half of the initial 
mass reservoir is left in the envelope, at $t\approx0.11$~Myr. 
The disk radius increases with time but also shows significant radial pulsations. These contractions/expansions
are caused by migration of the fragments and the corresponding redistribution 
of angular momentum within the disk, when the fragment's angular momentum is transferred to 
the spiral arms causing them to unwind and expand radially outward.
Overall, the considered model produces a rather massive and extended disk. Theoretically, such disks
are not unexpected in the embedded stage of star formation characterized by high rates of mass
infall onto the young disk. However, it must be kept in mind that we have not 
taken into account magnetic fields and the effect of the external environment, which
can significantly decrease the mass and size of protostellar disks 
\citep[see e.g.][]{HT08}. Therefore, although the qualitative picture issued from the present calculations should be rather robust, the quantitative results must be considered with due caution.

\begin{figure*}
 \centering
  \includegraphics[width=15cm]{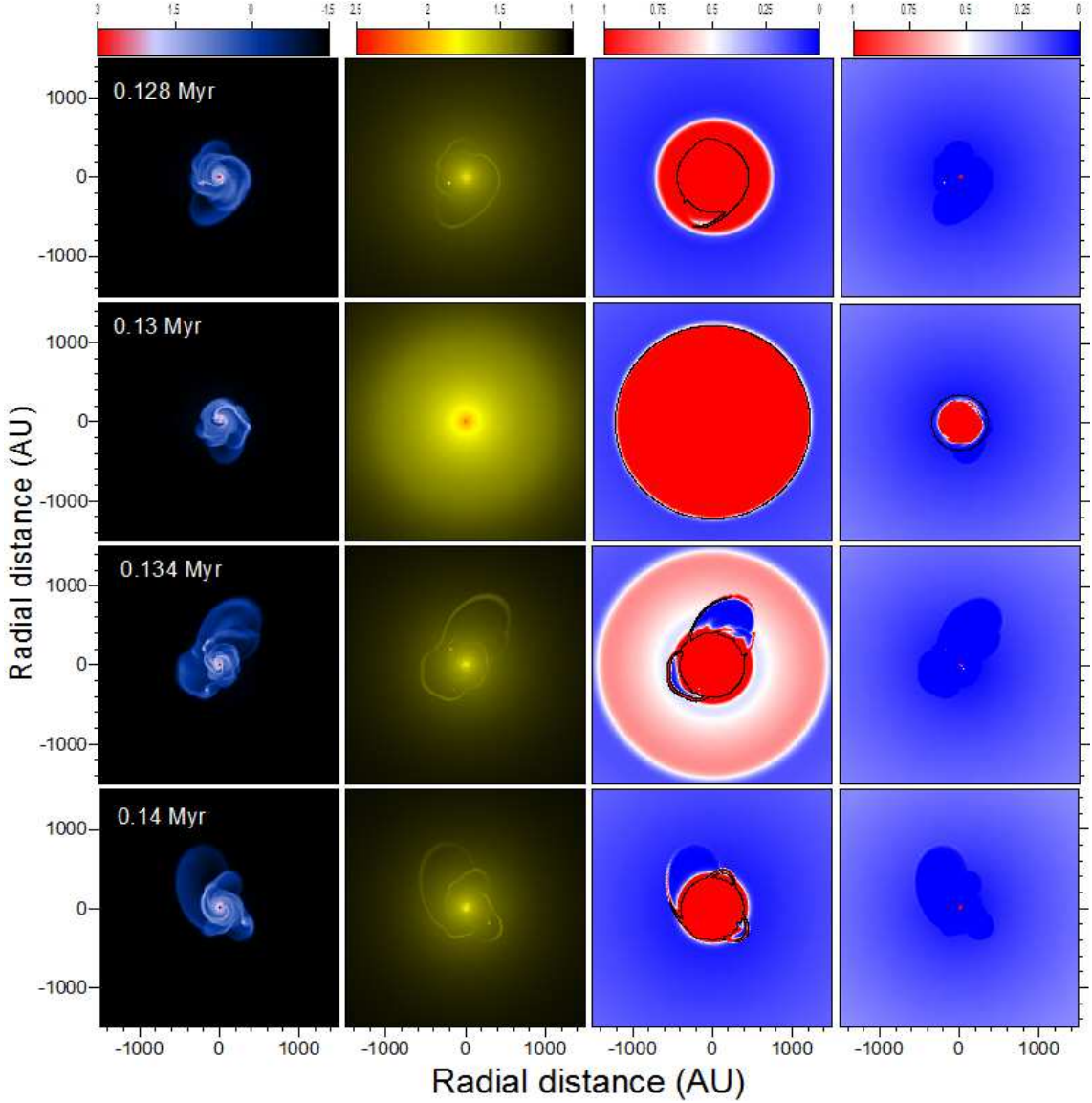}
  \caption{Effect of the luminosity bursts on the spatial distribution of gas-phase 
  CO and CO$_2$. The four columns present (from left to right) the gas surface density 
  distribution (log g cm$^{-2}$), gas temperature (K), CO gas-phase fraction, and CO$_2$ 
  gas-phase fraction in the inner 
  $3000\times3000$~AU region. The four rows show (from top to bottom) the pre-burst phase (t=0.128~Myr),
  the burst phase (t=0.13~Myr), the post-burst phase (t=0.134~Myr) and the quiescent phase 
  (t=0.14~Myr). The black lines are isotemperature contours outlining the regions where gas temperature
  exceeds 20~K (third column) and 40~K (fourth column) above which CO and CO$_2$ are supposed 
  to be in the gas phase in an equilibrium, time-independent case. }
  \label{fig3}
\end{figure*}

\subsection{Spatial distribution of CO and CO$_2$}
Figure~\ref{fig3} presents (from left to right) the gas surface density $\Sigma$ (first column), 
the gas temperature $T_{\rm mp}$ (second column), the CO gas-phase fraction
$\xi^{\rm g}_{\rm CO}=\Sigma_{\rm CO}^{\rm g}/(\Sigma_{\rm CO}^{\rm g} + \Sigma_{\rm CO}^{\rm s})$ (third
column) and the CO$_2$ gas-phase fraction $\xi^{\rm g}_{\rm CO_2}=\Sigma_{\rm CO_2}^{\rm g}/(\Sigma_{\rm CO_2}^{\rm g} + \Sigma_{\rm CO_2}^{\rm s})$ (fourth column) in the inner $3000 \times 3000$~AU region,
at four distinct time instances. 
The first row (from top to bottom) highlights the evolution stage
soon after a moderate luminosity burst  with $L_{\ast}\approx 20~L_\odot$ 
($\dot{M}\approx10^{-5}~M_\odot$~yr$^{-1}$) that occurred at $t=0.127$~Myr.
The second row presents the model during a strong luminosity burst with 
$L_{\ast}\approx 250~L_\odot$ ($\dot{M}\approx 2\times 10^{-4}~M_\odot$~yr$^{-1}$) 
at $t=0.13$~Myr. The third row shows the model at t=0.134~Myr, i.e., 4.0~kyr after the strong burst.
Finally, the fourth row represents the quiescent stage at $t=0.14$~Myr with $L_\ast \approx3.5~L_\odot$.
The black lines delineate isotemperature contours with  $T_{\rm mp}=20$~K (third column)
and $T_{\rm mp}=40$~K (fourth column). These values correspond to the evaporation temperatures
of CO and CO$_2$ ices from dust grains \citep{Noble12}.

A visual inspection of Figure~\ref{fig3} indicates that the protostellar disk (localized in the inner
300-350~AU) is characterized 
by almost complete evaporation of CO from dust grains ($\xi^{\rm g}_{\rm CO}\approx 1.0$), 
a result consistent with the recent study of molecular abundances in gravitationally
unstable disks by \citet{Ilee11}. This holds for both the burst and quiescent stages of disk evolution.
A moderate burst at $t=0.127$~Myr ($L_\ast\approx 20~L_\ast$) can also evaporate CO 
in the innermost parts of the envelope up to $r\approx500$~AU (first row) 
and the CO evaporation region can extend beyond 1000~AU for a strong burst with $L_\ast\approx250 L_\odot$
(second row). 

The burst duration is usually limited to 100--200~yr
(defined as the period of time during which the mass accretion rate constantly exceeds $10^{-5}~M_\odot$~yr$^{-1}$) but the effect of the burst is lingering in the envelope for a significantly longer time. 
For instance, the duration of
the burst at $t=0.13$~Myr is about 0.2~kyr, but a significant amount of gas-phase CO is seen in the
envelope at $t=0.134$~Myr, i.e., almost 4.0~kyr after the burst (third row in Figure~\ref{fig3}).  
The gas temperature in the envelope
at this stage drops below 20~K (as indicated by the isotemperature contour), which means that
the presence of gas-phase CO is truly related to a recent luminosity burst. 
Two lobe-like features that are almost completely devoid of gas-phase CO are evident in the third
row. The spiral arms have dynamically swept through the lobe regions
prior to the snapshot, and during this passage the CO gas has frozen out onto the grains in
the high-density wake of the arms. At the same time, the leading edge of the arms 
is shock compressed to a sufficiently high temperature  to evaporate CO.
Finally, we note that CO freezes out onto dust grains during the quiescent stage if its duration is
comparable to or longer than 10~kyr (fourth row in Figure~\ref{fig3}).

The effect of a recent luminosity burst on the gas-phase CO in the envelope can be
understood from the following simple analysis.  The $e$-folding time of freeze-out
onto dust grains $t_{\rm ads}=\lambda^{-1}$ can be expressed as
\begin{equation}
t_{\rm ads}= 7.7\times10^{-12} {a \over \rho_{\rm g} A_{\rm d2g} } \left( M_{\rm i} \over 
T_{\rm mp} \right)^{0.5}  \,\, [\mathrm{yr}],
\label{efold}
\end{equation} 
where $\rho_{\rm g}=\Sigma/2 h$ is the gas volume density. For the 
typical conditions in the inner envelope at $r=1000$~AU ($T_{\rm mp}=15$~K and
$n_{\rm g}=10^6$~cm$^{-3}$), the resulting e-folding time for CO freeze-out onto dust
grains is approximately 2500~yr. The fact that $t_{\rm ads}$ for carbon monoxide 
can be much longer than the burst duration
opens up a possibility for the observational
detection of recent bursts, as suggested by \citet{Lee07} and \citet{Visser12} on the basis of simplified
core collapse calculations. Equation~(\ref{efold}) also indicates that $t_{\rm ads}$ increases
linearly with the radius of dust grains $a$, suggesting that this phenomenon may become even 
more pronounced if dust
grains have enough time to coagulate and grow to sizes greater than adopted in the present study, $a=0.1\mu$m.

In contrast, the phase transformations of CO$_2$ during the bursts are much less pronounced.
Most of the disk and 
all of the envelope in Figure~\ref{fig3} are characterized by CO$_2$ frozen out onto grains.
CO$_2$ has an evaporation temperature of 35--40~K and in the quiescent stage
the gas-phase CO$_2$ is present only in the inner 25--30~AU. During the strong burst with $L_\ast\approx
250~L_\odot$ (second row in Figure~\ref{fig3}), the temperature
may rise above the CO$_2$ evaporation limit  in the inner 200--300~AU, transforming
most of CO$_2$ into the gas phase. However, the gas density in this
region of the disk ( $n_{\rm g}=(1-5) \times 10^{9}$~cm$^{-3}$) is considerably higher  
than in the envelope and the $e$-folding time for the CO$_2$ freeze-out onto dust grains 
($t_{\rm ads}\sim 1.0$~yr) is significantly shorter than that of carbon monoxide 
($t_{\rm ads}\sim 2500$~yr). As a result, the gas-phase CO$_2$ quickly returns into the solid phase
after the burst and no gas-phase CO$_2$ is seen at $t=0.134$~Myr at radial distances beyond 25--30 AU.
We conclude that the phase transitions of CO$_2$, and in particular the abundance of gas-phase
CO$_2$, are less convenient for monitoring the recent burst activity. 
We note, however, that the abundance of {\it solid} CO$_2$ appears
to be sensitive to the past accretion history and
can be a good episodic accretion tracer, as recently demonstrated by \citet{Kim11}.


\begin{figure*}
  \centering
  \includegraphics[width=15cm]{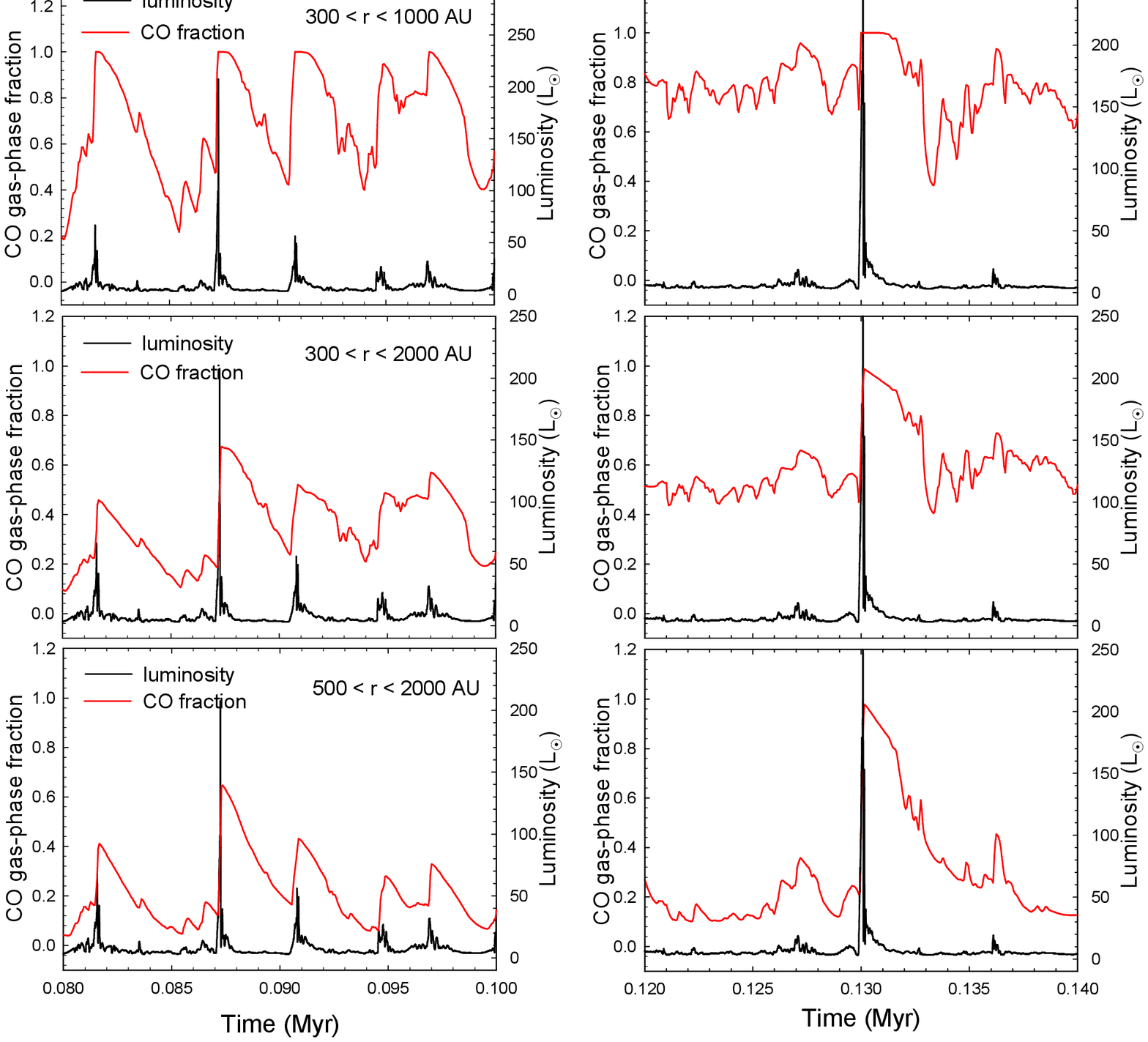}
  \caption{CO gas-phase fraction $\xi_{\rm CO}^{\rm g}$ (red lines) and total 
  stellar luminosity $L_\ast$ (black lines) vs. time.
  Two time periods (20~kyr each) are shown to highlight two strongest luminosity bursts
  and several moderate ones. The correlation between $\xi_{\rm CO}^{\rm g}$ and $L_\ast$ is evident.
  In particular, $\xi_{\rm CO}^{\rm g}$ steeply rises during the burst
  to a maximum value and gradually declines to a minimum value after the burst.}
  \label{fig4}
\end{figure*}

\subsection{Time variations of the gas-phase CO and CO$_2$ in the envelope}
In Figure~\ref{fig4} we consider the time evolution of the gas-phase CO fraction 
($\xi_{\rm CO}^{\rm g}$) during two time intervals: 0.08--0.1~Myr (left column) and
0.12--0.14~Myr (right column). These time intervals were chosen so as to capture two most 
intense and several moderate luminosity bursts. The red lines present $\xi_{\rm CO}^{\rm g}$ averaged
over three radial bins: $300<r<1000$~AU (top row), $300<r<2000$~AU (middle row), 
and $500<r<2000$~AU (bottom row).  The first
two bins cover the outer parts of the disk and the inner parts of the infalling envelope, while
the last bin covers only the inner envelope (the disk radius is less than 500~AU, see Figure~\ref{fig2}).
The time is counted from the formation of the central protostar.
The black lines show the total protostellar luminosity $L_\ast$ vs. time. 

The general correlation between $L_\ast$ and $\xi_{\rm CO}^{\rm g}$ 
is evident in Figure~\ref{fig4}. The CO gas-phase fraction steeply rises during the burst
to a maximum value and gradually declines to a minimum value after the burst. 
The relaxation time to the pre-burst stage 
is notably longer than the burst duration, in agreement with analytic estimates performed in 
the previous section. This pattern of behaviour---periodic steep rises 
during the bursts followed by gradual declines---is most pronounced in the envelope. 
Carbon monoxide in the disk is  mostly in the gas phase  and the bursts have little effect
on the CO gas-phase fraction there. The top-right panel in Figure~\ref{fig4} illustrates such an
example. The CO gas-phase fraction is averaged over the $300<r<1000$~AU radial bin, which 
covers part of the disk (with radius 350--400~AU at this time instant). Evidently, the correlation
between $\xi_{\rm CO}^{\rm g}$ and $L_\ast$ is less pronounced. It is therefore 
important to filter out the disk contribution when calculating $\xi_{\rm CO}^{\rm g}$.

\begin{figure}
  \resizebox{\hsize}{!}{\includegraphics{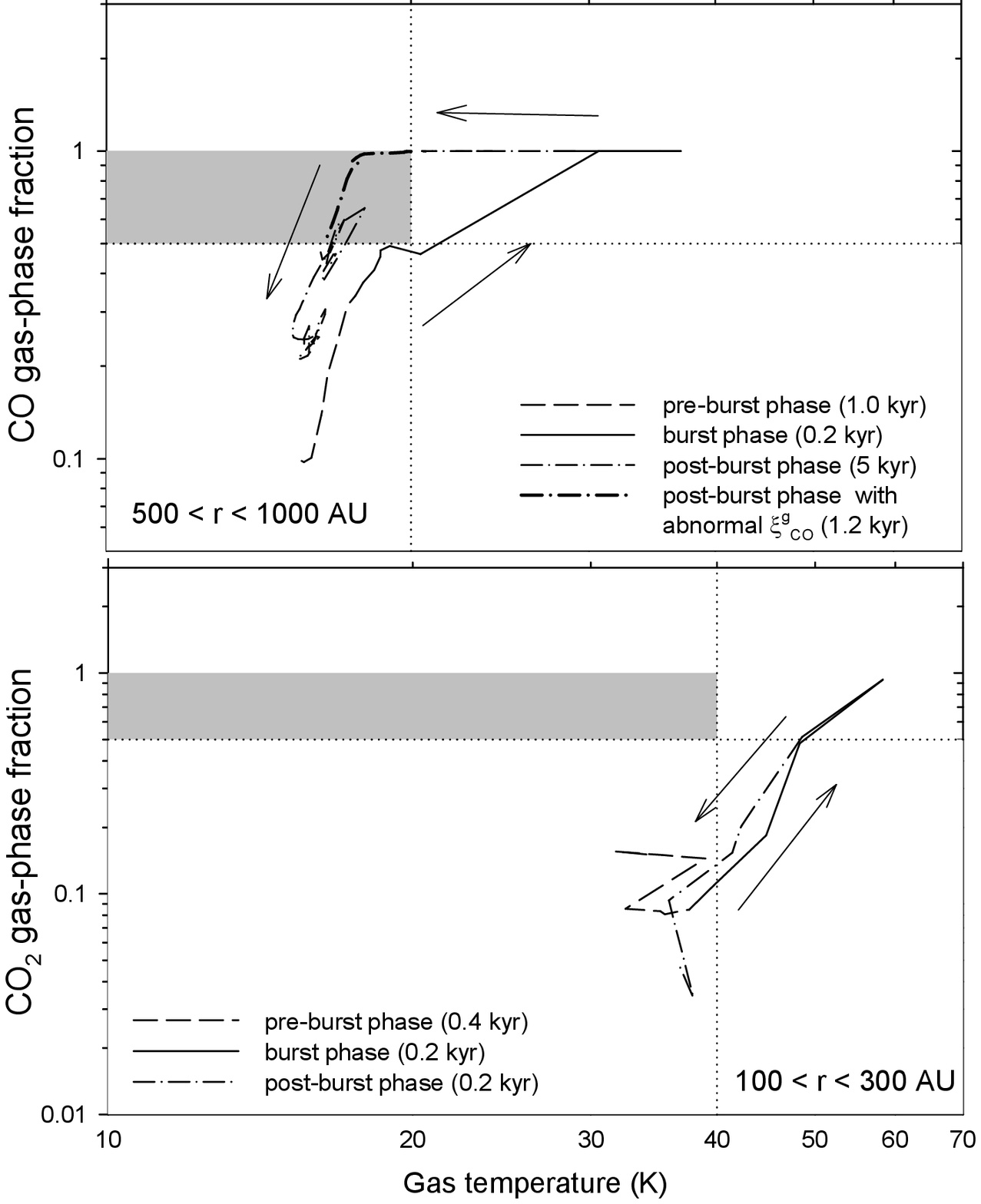}}
  \caption{{\bf Top.} CO gas-phase fraction $\xi_{\rm CO}^{\rm g}$ vs. gas temperature $T_{\rm mp}$
  calculated in the $500<r<1000$~AU radial annulus during a time period covering a strong luminosity
  burst at $t=0.13$~Myr. The dashed, solid, and dash-dotted lines show the pre-burst, burst, and post-burst
  phases with durations of 1.0~kyr, 0.2~kyr, and 5~kyr, respectively. The arrows indicate this evolution
  sequence. The horizontal and vertical
  dotted lines are the CO fraction of 0.5 and the gas temperature at
  which the CO desorbs. The grey-shaded area highlights the phase-space region with an
  abnormally high fraction of the gas-phase CO. 
  The thick dash-dotted line highlights the $\xi_{\rm CO}^{\rm g}$--$T_{\rm mp}$ model 
  track in this region.
  {\bf Bottom.} CO$_2$ gas-phase fraction $\xi_{\rm CO_2}^{\rm g}$ vs. $T_{\rm mp}$
  calculated in the $100<r<300$~AU radial annulus during a strong luminosity
  burst at $t=0.082$~Myr.  The dashed, solid, and dash-dotted lines show the pre-burst, burst, and post-burst
  phases with durations of 0.4~kyr, 0.2~kyr, and 0.2~kyr, respectively. The horizontal and vertical
  dotted lines mark the CO$_2$ fraction of 0.5 and the gas temperature at
  which the CO$_2$ desorbs. The grey-shaded area highlights the phase-space region with an
  abnormally high fraction of the gas-phase CO$_2$. }
  \label{fig5}
\end{figure}

The delayed adsorption of CO onto dust grains after a strong luminosity burst
is illustrated in the top panel of Figure~\ref{fig5} showing the gas-phase CO fraction 
$\xi_{\rm CO}^{\rm g}$ vs. gas temperature $T_{\rm mp}$ in the $500<r<1000$~AU radial annulus.
A short time period covering the burst at $t=0.13$~Myr is shown including 
1.0~kyr immediately before the burst (dashed line), 0.2~kyr during the burst (solid line), 
and 5~kyr after the burst (dash-dotted line). 
The vertical and horizontal dotted lines mark the CO fraction of 0.5 and the gas temperature above
which CO evaporates. It is evident that the evolution tracks of the CO gas-phase fraction are 
different in the pre- and post-burst phases. Before the burst CO is found mostly in the solid 
phase, as expected for $T_{\rm mp}<20$~K.
During the burst, $\xi_{\rm CO}^{\rm g}$ increases with the rising gas temperature and reaches a maximum
value of $\approx1.0$. However, after the burst the gas temperature
quickly drops below 20~K, but $\xi_{\rm CO}^{\rm g}$ remains abnormally high. 
The corresponding region in the $\xi_{\rm CO}^{\rm g}-T_{\rm mp}$ diagram is filled with a grey shade
and the model track in this region is highlighted by the thick dash-dotted line. The time spent
in this abnormal region (with $\xi_{\rm CO}^{\rm g}>0.5$ and $T_{\rm mp}<20$~K)
is 1.2~kyr\footnote{This value does not take into account the time spent at 
$\xi_{\rm CO}^{\rm g}>0.5$ and $T_{\rm mp}>20$~K. The full time needed for the gas-phase fraction 
of CO to go back down to 0.5 following the burst is about 2--3~kyr. }
The resulting mismatch between the gas temperature and the CO gas-phase content {\it in the post-burst phase} can be used to detect recent luminosity outbursts.  Objects that are found in the grey-shaded
area of the diagram are likely to be in the post-burst phase; objects in the quiescent 
and pre-burst phases are unlikely to fall into this region.


On the contrary, the phase transformation of CO$_2$ lacks such a characteristic feature.
The bottom panel in Figure~\ref{fig5} presents the gas-phase CO$_2$ fraction 
$\xi_{\rm CO_2}^{\rm g}$ vs. gas temperature $T_{\rm mp}$ in the $100<r<300$~AU radial annulus
during a short time period covering the burst at $t=0.082$~Myr. 
The dotted vertical line marks the critical gas temperature
above/below which CO$_2$ is supposed to be mostly in the gas/solid phase.
During the burst 
$\xi_{\rm CO_2}^{\rm g}$ rises to a maximum value of $\approx 1.0$ and returns to
a small value of $\le 0.1$ during just 0.2~kyr in the post-burst phase. 
There is {\it no} evolution stage when the gas temperature is below the condensation 
temperature of CO$_2$, but the gas-phase CO$_2$ fraction is abnormally high. The evolution 
tracks of the CO$_2$ gas-phase fraction in the pre- and post-burst phases follow a similar path 
in the $T_{\rm mp}-\xi_{\rm CO_2}^{\rm g}$ diagram and do not pass through the shaded region.
This makes CO$_2$ ineffective in detecting recent luminosity
bursts. 
 
\section{Model uncertainties}
\label{caveats}

Our model has a few assumptions that may affect the abundance of CO and CO$_2$ ices
in the disk and envelope. In this section, we discuss the model caveats and their impact  on our main results. 

{\it Uncertainties in the gas/dust temperature.} In our model we assumed that the gas and dust
temperatures are equal. This is usually correct for the gas densities typical for 
a protostellar disk 
($>10^{9}-10^{10}$~cm$^{-3}$), where frequent collisions between the gas and dust particles 
lead to fast thermalization between these species. In the envelope, however, the gas temperature
may differ from that of the dust. A more accurate approach with a separate treatment of dust and 
gas thermodynamics is needed to assess the possible effect of the gas to dust temperature 
imbalance. 

{\it The lack of vertical structure.} The use of the thin-disk approximation implies that
all quantities are averaged over the scale height and there are no variations in
the gas volume density and temperature with the distance from the midplane.
This may affect the adsorption and desorption rates of ices if strong vertical variations
in the thermodynamical properties are present. Work is in progress to reconstruct 
the vertical structure based on the coupled solution
of the radiation transfer and hydrostatic equilibrium equations. Preliminary results
indicate that notable vertical variations in the gas temperature and density may be present
in the inner disk regions but they are diminishing in the envelope 
(Vorobyov et al. 2013, in prep.). 

{\it Simplified treatment of gas thermodynamics.} Present results are based on a simplified 
form of diffusion approximation to calculate the thermal balance in the disk and envelope. 
We test our approach by comparing our derived gas/dust temperatures with those obtained with  an improved treatment of radiative transfer using the TORUS code \citep{Harries04,Harries11}. 
The later employs the Monte-Carlo algorithm described by \citet{Lucy99}. We
assume silicate dust grains with a dust-to-gas ratio of 0.01, and
adopt a canonical ISM grain-size distribution \citep{Mathis77}. 
The grain opacities and Mie phase matrices were
calculated from the refractive indices of astronomical silicates
(Draine and Lee 1984). Note that TORUS 
also assumes that dust and gas temperatures are the same.
 
\begin{figure}
  \resizebox{\hsize}{!}{\includegraphics{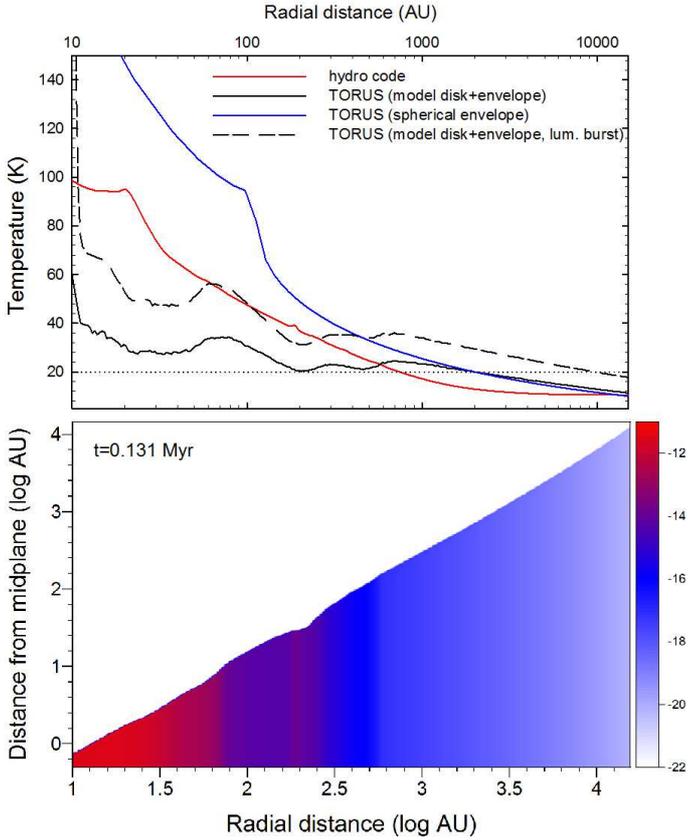}}
  \caption{{\bf Top.} Gas radial temperature profiles. In particular, the red line
  presents the azimuthally averaged midplane temperature $T_{\rm mp}$ 
  derived from hydrodynamical simulations at t=0.131~Myr and the total stellar luminosity 
  $L_\ast=9~L_\odot$. The black solid line shows the gas temperature
  derived using the TORUS radiation transfer code, for the same gas density distribution 
  and stellar luminosity as in the hydro model (case 1). The dashed solid line presents the TORUS calculation
  of the gas temperature for the same gas density distribution (case 1) but a higher stellar luminosity of 
  $100~L_\odot$. Finally, the blue line shows the gas temperature calculated
by the TORUS code for a spherically symmetric gas density distribution representing
an infalling spherical envelope with the same mass as in the hydro model (case 2).
  {\bf Bottom}. The gas volume density distribution used by the TORUS code (case 1) to calculate
  the gas temperatures (black solid and dashed lines). The scale bar is in g~cm$^{-3}$.} 
  \label{fig6}
\end{figure}

The post-processed radiative transfer calculations with TORUS require the knowledge 
of the density distribution. We explore two different cases with axisymmetric  (case 1) and spherically symmetric (case 2) distributions, respectively. Case 1 provides information on the temperature uncertainty resulting from the simplified treatment of radiative transfer in the hydrodynamical model. Case 2 corresponds to a spherically symmetric envelope and provides a limiting case which helps estimating the temperature
uncertainty in the envelope due to imposed thin-disk limit
(see below and Fig.~\ref{fig6}). The combination of both cases can thus provides a rough estimate of the error on temperatures determined in 
the hydrodynamical model. 

The top panel in Figure~\ref{fig6} presents the gas temperature vs. 
radial distance calculated using both our hydrodynamical
model (providing the so-called midplane gas temperature $T_{\rm mp}$) 
and the TORUS code. In particular, the red line shows the azimuthally averaged $T_{\rm mp}$
in our model in the post-burst stage at $t=0.131$~Myr when the luminosity of the central protostar has dropped to approximately
9.0~$L_\odot$. The solid black line (case 1) is the gas midplane temperature 
calculated using the TORUS code for the same gas density 
distribution and the same stellar luminosity as in our hydrodynamical model at $t=0.131$~Myr. 
A mass-weighted vertical mean temperature  differs from the midplane temperature in the TORUS calculation
by a factor of order unity. 
We convert the model surface densities into volume densities (needed in TORUS) 
using the following simple formula: $\rho(r)=\tilde{\Sigma}(r)/2\tilde{h}(r)$,
where $\tilde\Sigma$ and $\tilde{h}$ are the azimuthally averaged gas surface density and vertical 
scale height. The resulting gas density distribution $\rho(r)$ 
has a flared form as shown in the bottom panel of Figure~\ref{fig6}, with the vertical scale 
height increasing with distance. The color shaded area represents the disk plus envelope, 
while the white area is filled with a rarefied gas representing an outflow cavity.

The dashed black line (case 1) in top panel of Fig.~\ref{fig6} corresponds to the TORUS calculation of the gas temperature for the same model 
gas density distribution but for a higher luminosity of the central source during the burst, namely
$L_\ast=100~L_\odot$. 
A comparison between those different temperatures is mostly relevant 
for the envelope (i.e $r \simgt 300$ AU), since the radiative equilibrium calculations with 
TORUS only include heating from the central protostar and no dynamical effects of the gas 
(e.g viscous heating or compression), which are more important in the disk than in the envelope.  
This explains the significantly lower temperature obtained in the inner region ($r \simlt 300$ AU) 
with TORUS compared to $T_{\rm mp}$, for similar gas density distribution and protostar 
luminosity. The coupling of the hydrodynamical model with TORUS, providing a more consistent description of heating and cooling processes, is left for future work. 

In the outer regions ($r \simgt 300$ AU) the temperature differences between the TORUS temperature (case 1, solid black line)  and $T_{\rm mp}$ is less than 30\%. Though small, such temperature uncertainty may  be relevant if temperatures are close to condensation temperatures, as indicated by the horizontal dotted line for CO condensation in the upper panel of Fig.~\ref{fig6}. Reassuringly enough,  case 1 with a higher protostar luminosity (dashed black line) shows  the overall temperature increase expected during a burst.
Such a temperature increase will push
regions where CO may condense much further out ($>>$ 1000 AU), giving some confidence
in the abundance estimate of the gas-phase CO in the envelope derived in \S 2.3. Moreover, an overall decrease in the gas density with radial distance implies
that the typical freeze-out time of gas-phase CO onto dust grains will increase at larger distances,
enabling the detection of earlier bursts.

 Finally, the blue line in the top panel of Fig.~\ref{fig6} shows the gas temperature calculated
by the TORUS code for a spherically symmetric gas density distribution (case 2) representing
an infalling spherical envelope with mass equal to the envelope mass in our hydrodynamic model 
($0.5~M_\odot$). The gas density profile
in the envelope follows the relation $\rho(r) \propto r^{-2}$, characteristic for 
collapsing truncated cores of finite size \citep{VB05}.
The inner and outer radii of the envelope are set to 100~AU and 15000~AU.
The temperature difference resulting from different envelope geometries (TORUS case 1 versus TORUS case 2) can be large at $\sim$ 100 AU (about a factor three), but rapidly decreases as a function of radial distance. A spherical envelope is an extreme case, since outflows are expected to create a cavity which can extend as far as several thousand AU or more \citep{Com10,Machida2011b}. The spherical case is however interesting since it provides an upper limit to the expected temperature in the envelope. Current assumptions made in the hydrodynamical model are thus not expected to alter the conclusions derived from  the predicted gas-phase abundances of CO during and after a burst (\S 2.3). But the spherical case shows that taking into account the vertical structure of the disk and the envelope is important for a better estimate of the temperature.

{\it The lack of chemical reactions}.  In the present work, we neglect reactions that
can alter the chemical composition of protostellar disks and collapsing envelopes. In particular,
\citet{Kim11} suggest that CO turns into CO$_2$ during the quiescent accretion stage when CO freezes
out onto dust grains.  They found that up to 80\% of the original CO content can be converted into 
CO$_2$ in the envelope during the main accretion phase. 
We note, however, that the abundance of the gas-phase CO in the quiescent stage is non-negligible (see
Figure~\ref{fig4}), owing to a long characteristic freeze-out time, and 
this may reduce the efficiency of CO to CO$_2$ conversion on the surface of dust grains.
Detailed calculations using the grain-surface reactions are needed to
estimate the CO depletion due to this effect.

{\it The hybrid cold/hot accretion scheme.} In this study, we assumed that a fraction of 
the accretion energy is absorbed by the protostar when the accretion rate $\dot{M}$
exceeds a critical value of $\dot{M}_{\rm cr}=10^{-5}~M_\odot$~yr$^{-1}$. Below this value,
accretion onto the protostar proceeds in the cold regime, with all accretion energy radiated away.
The choice of $\dot{M}_{\rm cr}$ was motivated by fitting of the spectral energy distribution 
of FU Orionis by \citet{Hartmann2011}, who argued that these eruptive stars undergo 
expansion during the burst, which in turn indicates that some accretion energy is 
absorbed by the star. Accretion rates during FU-Orionis-type bursts typically lie
in the $10^{-6}$--$10^{-4}~M_\odot$~yr$^{-1}$ range.  
Varying the value of $\dot{M}_{\rm cr}$
in these limits, will affect the stellar radius $R_\ast$  and the resulting accretion and
photospheric luminosities. However,  the total stellar luminosity is not expected to vary significantly
because the increase in the accretion luminosity due to, say, decrease in the stellar radius, 
will be at least
partly compensated by the corresponding decrease in the photospheric luminosity. 



\section{Conclusions}

In this paper, we studied the phase transformations of CO and CO$_2$
during the early stages of low-mass star formation characterized by strong luminosity outbursts 
 resulting from accretion bursts.
We used numerical hydrodynamics modeling in the thin-disk limit to describe the gravitational 
collapse of a rotating pre-stellar core, extending our calculations into the disk formation 
stage and terminating the calculations when most of the initial pre-stellar core has accreted onto the disk 
plus protostar system. In the early accretion phase, the system experiences repetitive 
accretion and luminosity bursts caused by disk gravitational fragmentation and 
quick inward migration of the fragments onto the protostar \citep{VB10}.
The basic hydrodynamics equations were complemented with the continuity equations 
describing the adsorption and desorption of CO and CO$_2$ onto and from dust grains. 
We calculated the gas-phase fractions of CO and CO$_2$ ($\xi_{\rm CO}^{\rm g}$ and 
$\xi_{\rm CO_2}^{\rm g}$) in the protostellar disk and infalling envelope in the 
pre-burst, burst, and post-burst phases. We found the following.
\begin{itemize}
\item In the quiescent phase characterized by total stellar luminosity of the order of 
a few $L_\odot$, CO in the disk is found mostly in the gas phase, 
while in the envelope CO has mostly frozen
out onto dust grains. CO$_2$ is found in the gas-phase only in the inner 20--30~AU, 
while in the rest of the disk and envelope CO$_2$ freezes out onto dust.
\item During strong luminosity bursts characterized by a total luminosity from a few tens 
to a few hundreds $L_\odot$, CO ice evaporates from dust grains in part of the envelope, 
while CO$_2$ ice turns into the gas phase only in the inner several hundred AU of the disk.
\item The typical time for freeze-out of the gas-phase CO onto dust grains in the envelope 
(a few kyr) is considerably longer than the typical duration of a luminosity burst (0.1--0.2~kyr).
As a result, a significant amount of the gas-phase CO can still
be present in the envelope long after the system has returned into the quiescent phase.
This phenomenon can be used to infer recent luminosity bursts with magnitudes typical of
 EX-Lupi and FU-Orionis-type outbursts, as suggested by recent semi-analytical studies
by \citet{Lee07} and \citet{Visser12}.
\item In contrast, the typical freeze-out time of the gas-phase CO$_2$ is comparable to
the burst duration, owing to significantly higher gas densities in the disk. 
We thus conclude that 
CO$_2$  is probably not  a good tracer for recent burst activity in young protostars.

\end{itemize}
\smallskip

Regarding uncertainties of the present model, the heating due to stellar irradiation is crucial to derive temperatures in the disk and the envelope. In this work, we tried  to estimate uncertainties resulting from approximate treatment of radiative transfer in the hydrodynamical model by comparing temperatures obtained from an improved treatment of irradiation effects based on a radiative transfer code. The results of this comparison indicate that the uncertainty on the predicted temperatures 
depends on the radial distance from the protostar and the structure of the envelope. However, this 
uncertainty does not alter our conclusions regarding the abundance of gas-phase CO in the envelope and the possibility to use it as a tracer of recent accretion burst activity. 
Our short-term plans to reduce the present
uncertainties involving  (i) the improvement of radiative transfer calculations, based on the coupling between the hydrodynamical model and a radiative transfer code, like e.g TORUS, (ii) the reconstruction of the vertical structure of the disk, in order to improve over the thin disk approximation and (iii) the implementation of a chemical model, which is  more challenging 
but would provide more robust predictions.  
Finally, an important weakness of our model is the absence of magnetic fields, which are 
expected to alter the properties of the collapse, the disk (e.g the size)  and the accretion 
process onto the protostar.   Inclusion of magnetic field, with the improvement above-mentioned, is a major challenge that will provide the state-of-the-art on a longer term. 
Three-dimensional non-ideal MHD simulations including 
radiation hydrodynamics are now underway to explore 
the second collapse (Masson et al. 2012). 
The work presented here already provides consistent predictions that can
be tested against observations. The increased sophistication on the MHD,
thermal  and chemical treatments we propose will further improve our
understanding of embedded phases and episodic accretion.

 \section{Acknowledgments}
The authors are thankful to Ruud Visser, the referee, for suggestions that helped to improve 
the manuscript and to Shu-ichiro Inutsuka for stimulating discussions.
This work is supported by Royal Society awards WM090065 and RFBR Cost shared application with Russia
(JP101297 and 11-02-92601).  
 It was also partly supported by the European Research Council under 
the European CommunityÕs Seventh Framework Programme (FP7/2007-2013 Grant Agreement No. 247060) and by the Consolidated STFC grant  ST/J001627/1. 
The simulations were performed
on the Shared Hierarchical Academic Research Computing Network (SHARCNET),
on the Atlantic Computational Excellence Network (ACEnet), and on the Vienna Scientific
Cluster (VSC-2).


\begin{thebibliography}{}

\bibitem[\protect\citeauthoryear{Armitage et al.}{2001}]{Armitage01}
Armitage, P. J., Livio, M., \& Pringle, J. E. 2001, MNRAS, 324, 705

\bibitem[\protect\citeauthoryear{Baraffe et al.}{2002}]{Baraffe02}
Baraffe, I., Chabrier, G., Allard, F. \& Hauschildt P. H. 2002, A\&A, 382, 563

\bibitem[\protect\citeauthoryear{Baraffe et al.}{2009}]{Baraffe09}
Baraffe, I., Chabrier, G., \& Gallardo, J. 2009, ApJ, 702, L27 

\bibitem[\protect\citeauthoryear{Baraffe \& Chabrier}{2010}]{Baraffe10}
Baraffe, I., \& Chabrier, G. 2010, A\&A, 521, 44 

\bibitem[\protect\citeauthoryear{Baraffe et al.}{2012}]{Baraffe12}
Baraffe, I., Vorobyov, E. I., \& Chabrier, G. 2012, ApJ, 756, 118

\bibitem[\protect\citeauthoryear{Basu \& Mouschovias}{1994}]{Basu94}
Basu, S., \& Mouschovias, T. Ch. 1994, ApJ, 432, 720

\bibitem[\protect\citeauthoryear{Basu}{1997}]{Basu97}
Basu S., 1997, ApJ, 485, 240

\bibitem[\protect\citeauthoryear{Bell \& Lin}{1991}]{Bell94}
Bell, K. R., \& Lin, D. N. C., 1994, ApJ, 427, 987

\bibitem[\protect\citeauthoryear{Bisschop et al.}{2006}]{Bisschop06}
Bisschop, S. E., Fraser, H. J., Oberg, K. I., van Dishoeck, E. F., \& Schlemmer, S. 2006, 
A\&A, 449, 1297



\bibitem[\protect\citeauthoryear{Bonnell \& Bastien}{1992}]{BB92}
Bonnell, I., \& Bastien, P. 1992, ApJ, 401, L31

\bibitem[\protect\citeauthoryear{Bourke et al.}{2006}]{Bourke06}
Bourke, T. L., Myers, P. C., Evans, N. J., II, et al. 2006, ApJ, 649, L37 

\bibitem[\protect\citeauthoryear{Chabrier \& Baraffe}{1997}]{Chabrier97}
Chabrier, G., \& Baraffe I. 1997, A\&A, 327, 1039

\bibitem[\protect\citeauthoryear{Chabrier \& Baraffe}{2000}]{Chabrier00}
Chabrier, G., \& Baraffe I. 2000, ARA\&A, 38, 337

\bibitem[\protect\citeauthoryear{Charnley et al.}{2001}]{Charnley01}
Charnley, S. B., Rodgers, S. D., \& Ehrenfreund, P. 2001, A\&A, 378, 1024

\bibitem[\protect\citeauthoryear{Commer\c con et al.}{2010}]{Com10}
Commer\c con, B., Hennebelle, P., Audit, E., Chabrier, G., \& Teyssier, R. 2010, A\&A, 510, 3

\bibitem[\protect\citeauthoryear{D'Angelo \& Spruit}{2010}]{DAS10}
D'Angelo, C. R., \& Spruit, H. C. 2010, MNRAS, 406, 1208

\bibitem[\protect\citeauthoryear{Dapp \& Basu}{2009}]{Dapp09}
Dapp, W. B., \& Basu, S. 2009, MNRAS, 395, 1092

\bibitem[\protect\citeauthoryear{Draine et al.}{1984}]{Draine84}
Draine, B. T., \& Lee, H. M. 1984, ApJ, 285, 89

\bibitem[\protect\citeauthoryear{Dunham et al.}{2006}]{Dunham06}
Dunham, M. M., Evans, N. J., II, Bourke, T. L., et al., 2006, ApJ, 65, 945

\bibitem[\protect\citeauthoryear{Dunham et al.}{2010}]{Dunham10}
Dunham, M. M., Evans II, N. J., Terebey, S., Dullemond, C. P., \& Young, C. H. 2010, ApJ, 710, 470

\bibitem[\protect\citeauthoryear{Dunham \& Vorobyov}{2012}]{DV12}
Dunham, M. M., \& Vorobyov, E. I. 2012, ApJ, 747, 52

\bibitem[\protect\citeauthoryear{Enoch et al.}{2009}]{Enoch09}
Enoch, M. L., Evans II, N. J., Sargent, A. I., \& Glenn, J. 2009, ApJ, 692, 973 

\bibitem[\protect\citeauthoryear{Evans et al.}{2009}]{Evans09}
Evans, N. J., II, Dunham, M. M., Jorgensen, J. K., et al., 2009, ApJSS, 181, 32

\bibitem[\protect\citeauthoryear{Harries et al.}{2004}]{Harries04}
Harries, T., Monnier, J. D., Symington, N., \& Kurosawa, R. 2004, MNRAS, 350, 565

\bibitem[\protect\citeauthoryear{Harries}{2011}]{Harries11}
Harries, T. 2011, MNRAS, 411, 1500

\bibitem[\protect\citeauthoryear{Hartmann et al.}{2011}]{Hartmann2011}
Hartmann L., Zhu Z., \& Calvet N., 2011, preprint (arXiv:1106.3343)

\bibitem[\protect\citeauthoryear{Hennebelle \& Teyssier}{2008}]{HT08}
Hennebelle, P., \& Teyssier, R. 2008, A\&A, 477, 25

\bibitem[\protect\citeauthoryear{Ilee et al.}{2011}]{Ilee11}
Ilee, J. D., Boley, A. C., Caselli, P., Durisen, R. H., Hartquist, T. W., \& Rawlings, J. M. C.
2011, MNRAS, 417, 2950

\bibitem[\protect\citeauthoryear{Johnson \& Gammie}{2003}]{Johnson03}
Johnson, B. M., \& Gammie C. F. 2003, ApJ, 597, 131

\bibitem[\protect\citeauthoryear{Kenyon et al.}{1990}]{Kenyon90}
Kenyon, S. J., Hartmann, L. W., Strom, K. M., \& Strom, S. E. 1990, ApJ, 99, 869 

\bibitem[\protect\citeauthoryear{Kim et al.}{2011}]{Kim11}
Kim, H. J., Evans, II, N. J., Dunham, M. M., et al. 2011, ApJ,
729, 84

\bibitem[\protect\citeauthoryear{Kim et al.}{2012}]{Kim12}
Kim, H. J., Evans, N. J., II, Dunham, M. M., Lee, J.-E., \& Pontoppidan, K. M.
2012, ApJ, 758, 38

\bibitem[\protect\citeauthoryear{Lee}{2007}]{Lee07}
Lee, J.-E. 2007, J. Korean Astron. Soc., 40, 83

\bibitem[\protect\citeauthoryear{Lin \& Papaloizou}{1986}]{Lin86}
Lin D. N. C., \& Papaloizou J. 1986, ApJ, 309, 846

\bibitem[\protect\citeauthoryear{Lodato \& Clarke}{2004}]{LC04}
Lodato, G., \& Clarke, C. J. 2004, MNRAS, 353, 841

\bibitem[\protect\citeauthoryear{Lucy}{1999}]{Lucy99}
Lucy, L. B. 1999, A\&A, 344, L282

\bibitem[\protect\citeauthoryear{Masson et al.}{2012}]{Masson12}
Masson, J., Teyssier, R., Mulet-Marquis, C., Hennebelle, P., \& Chabrier, G. 2012, ApJS, 201, 24

\bibitem[\protect\citeauthoryear{Machida et al.}{2011a}]{Machida2011}
Machida, M. N., Inutsuka, S., \& Matsumoto, T. 2011a, ApJ, 729, 42

\bibitem[\protect\citeauthoryear{Machida et al.}{2011b}]{Machida2011b}
Machida, M., Inutsuka, S-I., \& Matsumoto, T., 2011b, PASJ, 65, 555

\bibitem[\protect\citeauthoryear{Mathis et al.}{1977}]{Mathis77}
Mathis, J. S.,  Rumpl, W., \& Nordsieck, K. H. 1977, ApJ, 217, 425

\bibitem[\protect\citeauthoryear{McKee \&  Ostriker}{2012}]{mckee12}
McKee, C., \& Ostriker, ARA\&A, 45, 565




\bibitem[\protect\citeauthoryear{Noble et al.}{2012}]{Noble12}
Noble, J. A., Congiu, E., Dulieu, F., \& Fraser, H. J. 2012, MNRAS, 421, 768

\bibitem[\protect\citeauthoryear{Pfalzner et al.}{2008}]{Pfalzner08}
Pfalzner, S., Tackenberg, J., \& Steinhausen, M. 2008, A\&A, 487, L45

\bibitem[\protect\citeauthoryear{Pontoppidan et al.}{2008}]{Pontoppidan08}
Pontoppidan, K. M., Boogert, A. C. A., Fraser, H. J., et al. 2008, ApJ, 678, 1005

\bibitem[\protect\citeauthoryear{Shu}{1977}]{Shu77}
Shu, F. S. 1977, ApJ, 214, 488 

\bibitem[\protect\citeauthoryear{Stamatellos et al.}{2011}]{Stamatellos11}
Stamatellos, D., Whitworth, A. P., \& Hubber, D. A. 2011, ApJ, 730, 32

\bibitem[\protect\citeauthoryear{Takeuchi \& Lin}{2002}]{Takeuchi02}
Takeuchi, T., \& Lin, D. N. C. 2002, ApJ, 581, 134

\bibitem[\protect\citeauthoryear{Viallet \& Baraffe}{2012}]{Viallet12}
Viallet, M.,  \& Baraffe, I. 2012, A\&A, 546, 113


\bibitem[\protect\citeauthoryear{Visser et al.}{2009}]{Visser09}
Visser, R., van Dishoeck, E. F., Doty, S. D., \& Dullemond S. P. 2009, A\&A, 495,
881


\bibitem[\protect\citeauthoryear{Visser \& Bergin}{2012}]{Visser12}
Visser, R., \& Bergin, E. A. 2012, ApJ, 754, 18

\bibitem[\protect\citeauthoryear{Vorobyov \& Basu}{2005}]{VB05}
Vorobyov, E. I., \& Basu, S. 2005, MNRAS, 360, 675

\bibitem[\protect\citeauthoryear{Vorobyov \& Basu}{2006}]{VB06}
Vorobyov, E. I., \& Basu, S., 2006, ApJ, 650, 956

\bibitem[\protect\citeauthoryear{Vorobyov \& Basu}{2010}]{VB10}
Vorobyov, E. I., \& Basu, S. 2010, ApJ, 719, 1896

\bibitem[\protect\citeauthoryear{Vorobyov}{2012}]{Vor2012}
Vorobyov, E. I. 2012, Astron. Reports, 56, 179


\bibitem[\protect\citeauthoryear{Zhu et al.}{2010}]{Zhu10}
Zhu, Z., Hartmann, L., Gammie, C., Laura, G. B., Jacob, B. S., \& Eric, E. 2010,
ApJ, 713, 1134

\bibitem[\protect\citeauthoryear{Zhu et al.}{2012}]{Zhu2012}
Zhu, Z., Hartmann, L., Nelson, R. P., \& Gammie, C. F. 2012, ApJ, 746, 110












\end{thebibliography}
\end{document}